\documentclass[superscriptaddress,showkeys,preprintnumbers,amsmath,amssymb,floatfix,prd]{revtex4}

\usepackage{graphicx}
\usepackage{dcolumn}
\usepackage{bm}
\usepackage{hyperref}
\usepackage{orcidlink}
\hypersetup{colorlinks,citecolor=blue}
\begin{document}

\preprint{}

\title{Relativistic compact stars coupled with  dark energy in Heintzmann spacetime}

\author{Susmita Sarkar\orcidlink{0009-0007-1179-2495} }
\email{ susmita.mathju@gmail.com}
\affiliation{Department of Applied Science and Humanities, Haldia Institute of Technology, Haldia-721606, West Bengal, India}

\author{Nayan Sarkar\orcidlink{0000-0002-3489-6509} }
\email{ nayan.mathju@gmail.com}
\affiliation{Department of Mathematics, Karimpur Pannadevi College, Karimpur-741152, Nadia, West Bengal, India}

\date{\today}
\begin{abstract}
The literature suggests that dark energy is responsible for the accelerating expansion of the universe due to its negative pressure, therefore, dark energy can be used as a possible option to prevent the gravitational collapse of compact objects into singularities.  In this regard, there is a great possibility that dark energy can interact with the compact stellar matter configuration [Phys. Rev. D 103, 084042 (2021)]. In this article, we introduce a physically viable model for celestial compact stars made of isotropic baryonic matter and isotropic dark energy with Heintzmann's {\it ansatz} [Zeitschrift für Physik 228, 489-493 (1969)] in the context of Einstein's gravity.  Here, the density of dark energy is assumed to be proportional to the density of baryonic matter. The main focus of the present article is to see the effects of dark energy on the physical properties of the stars. We perform an in-depth analysis of the physical attributes of the model, such as metric function, density, pressure, mass-radius relation, compactness parameter, gravitational and surface redshifts, along with the energy conditions for three well-known compact stars. We analyse the equilibrium of the present model via the generalised Tolman-Oppenheimer-Volkoff equation and the stability with the help of the adiabatic index and Harrison-Zeldovich-Novikov’s static stability condition. Moreover, we estimate the solutions representing the maximum masses and the predicted surface radii from the $M-R$ graph for different values of the coupling parameter $\alpha$. All the analyses ensure that the present model is non-singular and physically viable by satisfying all the essential conditions.

\end{abstract}

\maketitle


\section{Introduction}\label{Sec1}

The structure of relativistic stars and their gravitational collapse have been key topics of great interest within the relativistic scientific community since the advent of general relativity. Indeed, the pioneering work of Schwarzschild \cite{ks16}, Fowler \cite{rh26}, Tolman \cite{rc39}, and Oppenheimer and Volkoff \cite{jr39} played a pivotal role in advancing the development of theoretical models for relativistic stars. Schwarzschild \cite{ks16} introduced the analytic solutions for the celestial compact objects with uniform density, Fowler \cite{rh26} first described that the Fermi-Dirac statistics is responsible for the high degeneracy pressure within the compact stars that holds up the stars against gravitational collapse,  Tolman \cite{rc39} devised a method that provides explicit solutions for static matter configurations, which has been instrumental in advancing the study of stellar structures, and Volkoff \cite{jr39} investigated the gravitational equilibrium of neutron stars for the specific Tolman solutions by applying the equation of state (EoS) of cold Fermi gas. In this regard,  Chandrasekhar \cite{sc31} studied the effects of special relativity on the EoS for a degenerate Fermi gas and demonstrated the existence of a maximum mass for compact stars.  Beyond this threshold, stars become unstable and undergo collapse, a phenomenon now referred to as the Chandrasekhar mass limit, $M_{ch} = 1.4 M_\odot$. The recent observational data obtained from the gravitational waves, particularly the GW170817 \cite{bp17} and GW190814 \cite{ra20} events, have reinvigorated the scientific community to explore the underlying dynamics and relativistic evolution of both gravitationally stable and unstable compact astrophysical objects \cite{zy18, rn18, gb19, da21}.

Several astrophysical and cosmological observations strongly suggest that the total mass of the universe is predominantly composed of non-baryonic mass and energy \cite{gb05, ad14}. The matter content of the universe comprises approximately 4.9$\%$ baryonic matter and 26.4$\%$ of an unseen form of matter known as dark matter (DM), whose existence is inferred from its gravitational effects and  68.7$\%$ dark energy (DE) with strong negative pressure \cite{pc16}. The various high-precision observational data, including galaxy rotation curves, Type Ia supernovae \cite{ag01, sp199}, and cosmic microwave background radiation \cite{cl03, gh03} confirmed the present accelerated expansion of the universe. While the attractive force of gravity would typically slow the expansion of the universe over time, the presence of dark energy explains the observed acceleration of that expansion \cite{na03, pj03, ej06, rr09, dn15, bw16}. As the dominant factor driving the universe's accelerated expansion, the study of dark energy, along with its significant effects on all forms of matter, energy, and space-time, has emerged as a vital area of research \cite{na20}. In this regard,  several researchers are adopting a new perspective that accounts for the interaction profile of dark energy in the overall evolution of compact structures, rather than neglecting it \cite{aa22, gb22, sc22, os22}. The concept of a dark energy star suggests that the surface of a compact object is a quantum-critical shell with a certain thickness, denoted by $z$ \cite{CG92}. It has been seen that ordinary elementary particles with energy exceeding $Q = 100MeV \sqrt{M_\odot/M}$, where $M_\odot$ is the solar mass, enter the quantum critical region and they decay into constituent particles and radiation that are emitted backwards from the surface of the dark energy star, perpendicular to the critical surface. However, particles with energy less than  $Q$ will pass through the critical surface and follow diverging geodesics within the interior of the dark energy star. It has been discovered that compact objects at the centre of galaxies contain quarks and gluons within nucleons whose energy exceeds the threshold value of $Q$ \cite{JB04}. Chapline et al. \cite{gc01} showed that the primordial dark energy stars may form from spacetime fluctuations, similar to a quantum critical instability. Indeed, the dark energy stars have been explored in numerous significant studies.  Lobo \cite{sn06} explored a model for stable dark energy stars by considering two different spatial mass functions: one based on a constant energy density and the other using the Tolman-Whitaker mass function.
Recently, Herrera \cite{lh20} confirmed that the unequal principal stresses in a gravitationally compact system cannot be ignored. In a relativistic framework, anisotropy becomes an inherent characteristic of the system, particularly in the hydrodynamic balancing equations. Recognising this intrinsic anisotropy, Sagar et al. \cite{kg22} explored the potential formation of stable compact stars by incorporating first-order coupling of the dark energy scalar field within the Tolman-Buchdahl spacetime. Indeed, their study revealed an intriguing result, showing that dark energy can offer quasi-dynamic stability to a compact structure within a gravitationally stable configuration. 

In recent years, several theoretical researchers have developed different models for the celestial compact stars coupled with dark energy. Saklany et al. \cite{ss22} studied the compact star model PSRJ1614-2230 coupled with dark energy in the framework of Tolman–Kuchowicz spacetime.  Sagar et al. \cite{kg22a} further investigated the first-order coupling of background dark energy with the hadronic matter for the compact star PSRJ0740+6620 using the embedding class-one approach, and interestingly, their study once again highlighted the role of dark energy as a quasi-dynamic stabilizer along with the existence of a forbidden region for the coupling parameter and the presence of a transition zone within the interior of the stellar structure. Bhar et al. \cite{pb15} proposed a novel dark energy star model composed of five distinct zones: a solid core with constant energy density, a thin shell between the core and the interior, an inhomogeneous interior region with anisotropic pressures, another thin shell, and an exterior vacuum region. Yadav et
al. \cite{ak11} studied a dark energy compact stars model by considering a variable
equation of state parameter. Yazadjiev \cite{ss11} introduced the exact interior solutions of the Einstein field equations describing mixed relativistic stars composed of ordinary matter and dark energy. Dayanandan et al.  \cite{bd21} proposed an anisotropic dark energy star model in the Tolman IV spacetime satisfying all energy conditions except the strong energy condition. Smerechynskyi et al. \cite{ss21} explored the density distribution of minimally-coupled scalar field dark energy within a neutron star. Indeed, their research examines how the presence of dark energy influences the star's macroscopic properties and impacts the upper mass limit, based on the predicted dark energy density distribution inside the star. Very recently, Rej et al. \cite{PR24} studied the Finch-Skea dark energy stars with the vanishing complexity factor. Besides, many different models of dark energy stars with several aspects have been introduced into the literature \cite{PB21, PB18, RB16, KD15, PH13, FR12, AB22, KP23}.

In this paper, we explore a new model for relativistic stars composed of both ordinary matter and dark energy in Heintzmann spacetime in the context of Einstein's gravity. Due to the presence of dark energy, it can be assumed that celestial compact stars consist of a mixture of ordinary matter and dark energy in different proportions. Indeed, the mixed composition of matter within stars has sparked significant interest within the scientific research community, and some more notable studies have already been made in this area of research \cite{RC09, AB23, CR11, SS12, DH13, PR23}. All these notable studies have motivated us to generalise a new model for dark energy compact stars with the Heintzmann metric potentials in Einstein's gravity. The present model has been analysed using some well-known compact stars, Vela X-1, Cen X-3, and SMC X-4. The compact star Vela X-1 is a high-mass X-ray binary (HMXB) system, consisting of a neutron star orbiting closely around the 23.5$M_\odot$B0.5Ib supergiant donor star HD 77581. Observations from a rocket-borne experiment\cite{GC67} and the Uhuru satellite \cite{RG72} revealed high variability in the matter source. The compact star Cen X-3, the most luminous X-ray pulsar in our galaxy, was first discovered using a rocket-borne detector in 1967 \cite{GC67}. Subsequently, satellite observations by Giacconi et al. \cite{RG71} and Schreier et al. \cite{ES72} revealed its binary and pulsar characteristics. The compact star SMC X-4 was detected by Price et al. \cite{RE71} in the Small Magellanic Cloud. It has been found that the compact star SMC X-4 has significant variability in both its intensity and spectrum \cite{CL71}. Schreier et al. \cite{ES72a} identified its binary nature and reported periodic occultations with an orbital period of approximately 3.9 days. In the year 1972, the Uhuru satellite first observed the compact star LMC X-4 \cite{RG72}, and later, Chevalier and Ilovaisky \cite{CC77} uncovered the binary nature of its optical counterpart. The structure of this article is organised as follows: We have described the interior spacetime with the Einstein field equations for a static and spherically symmetric matter sphere in Sec. \ref{Sec2}. In Sec. \ref{Sec3}, we have solved the field equations by considering the  Heintzmann spatial metric potential. The matching of our internal solution with the external Schwarzschild solution is established in Sec. \ref{Sec4} to determine the value of model constants.
  We have analysed several physical attributes of our proposed model in Sec. \ref{Sec5}. The equilibrium status of the present model is analyzed through the generalized TOV equation in Sec. \ref{Sec6}. In \ref{Sec7}, we have explored the stability of the present model and Sec. \ref{Sec8} deals with the maximum mass and the moment of inertia of the system. Finally, we have described the results and conclusion of this article in Sec. \ref{Sec9}.

\section{INTERIOR SPACETIME with Einstein's  field equations}\label{Sec2}

To describe the interior spacetime of a static, spherically symmetric celestial compact object, we consider  a 4D static, spherically symmetric line element in the Schwarzschild coordinate system $(t, r, \theta, \phi)$ as follows
\begin{eqnarray}
 ds^2=e^{\nu(r)} dt^2-e^{\lambda(r)} dr^2-r^2(d\theta^2+sin^2\theta d\phi^2).\label{ds}
\end{eqnarray}
Here, the metric potentials $e^{\nu(r)}$ and $e^{\lambda(r)}$ are assumed to be functions of ‘r’ only, reflecting the static nature of the spacetime. 

\begin{figure}[!htbp]
\begin{center}
\begin{tabular}{rl}
\includegraphics[width=8.8cm]{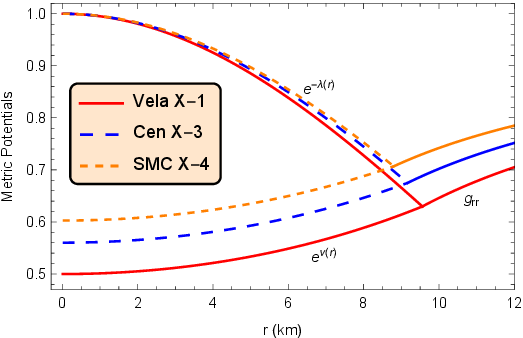}
\includegraphics[width=8.8cm]{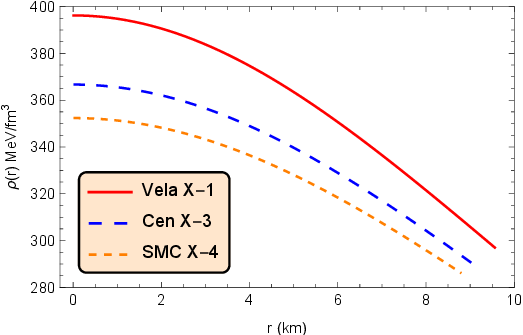}
\\
\end{tabular}
\end{center}
\caption{Profiles of metric potentials, Schwarzchild metric $g_{rr}(r)$ (Left) and energy density (Right) against the radial coordinate $r$ for the three model compact stars associated with the values of constants provided in Table-{\ref{tab1}}.}\label{fig1}
\end{figure}

The Einstein field equations can be written as (Gravitational unit, $G = c = 1$) 
\begin{eqnarray}
 R_{\mu\nu}-\frac{1}{2}g_{\mu\nu}R=8\pi T_{\mu\nu},\label{eq1}
\end{eqnarray}
where $R_{\mu\nu}$ is the Ricci tensor, $g_{\mu\nu}$ is the metric tensor, $R$ is the scalar curvature, and $T_{\mu\nu}$ is the energy momentum tensor. In this study, we consider that the energy-momentum tensor consists of the normal baryonic matter with energy density $\rho(r)$ and pressure $P(r)$, along with isotropic dark energy characterized by the energy density $\rho^{de}(r)$ and pressure $P^{de}(r)$. It is important to note that the dark energy density can be represented in terms of the cosmological constant as $\rho^{de}(r) =\frac{\Lambda}{8\pi}$ \cite{cr11}. Consequently, the corresponding non-zero components of the energy-momentum tensor of the two fluids can be expressed as \cite{cr11}
\begin{eqnarray}
T_0^0&=&\rho^{\text{eff}}(r)=\rho(r)+\rho^{de}(r),\label{eq3}
\\
T_1^1&=&T_2^2=T_3^3=-P^{\text{eff}}(r)=-[P(r)+P^{de}(r)].\label{eq4}
\end{eqnarray}
Here, $\rho^{\text{eff}}(r)$, and $P^{\text{eff}}(r)$ represent the effective energy density, and pressure, respectively. 

Now, the Einstein field equations (\ref{eq1}) for the line element (\ref{ds}) and energy-momentum tensor (\ref{eq3})-(\ref{eq4}) can be expressed as
\begin{eqnarray}
\rho(r)+\rho^{de}(r)&=&\frac{e^{-\lambda}}{8\pi}\left\{\frac{\lambda^\prime}{r}-\frac{1}{r^2}\right\}+\frac{1}{8\pi r^2}, \label{rhoeff}
\\
P(r)+ P^{de}(r)&=&\frac{e^{-\lambda}}{8\pi}\left\{\frac{\nu^\prime}{r}+\frac{1}{r^2}\right\}-\frac{1}{8\pi r^2}, \label{preff}
\end{eqnarray}
\begin{eqnarray}
P(r)+ P^{de}(r)&=&\frac{e^{-\lambda}}{32\pi}\left\{2\nu^{\prime\prime}+\nu^{\prime 2}+\frac{2(\nu^{\prime}-\lambda^{\prime})}{r}-\nu^{\prime}\lambda^{\prime}\right\}.\label{pteff}
\end{eqnarray}

Here, (') stands for the derivative with respect to the radial coordinate $r$. Now, we are going to find the physically acceptable solutions of the field equations (\ref{rhoeff})-(\ref{pteff}) in the background of the Heintzmann metric.

\begin{figure}[!htbp]
\begin{center}
\begin{tabular}{rl}
\includegraphics[width=8.8cm]{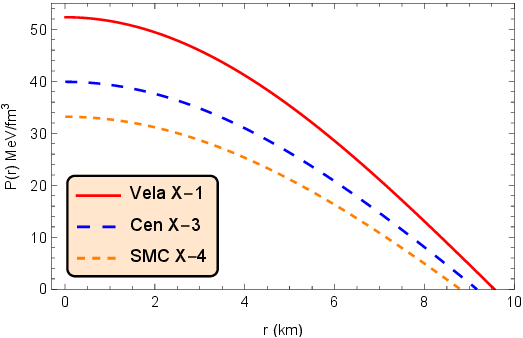}
\includegraphics[width=8.8cm]{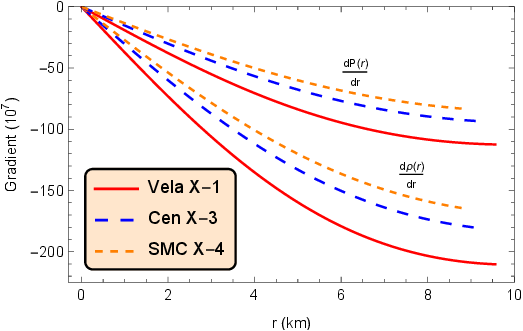}
\\
\end{tabular}
\end{center}
\caption{Profiles of pressure (Left) and gradients (Right) against the radial coordinate $r$ for the three model compact stars associated with the values of constants provided in Table-{\ref{tab1}}.}\label{fig2}
\end{figure}

\section{ Interior Solutions with Heintzmann metric  }\label{Sec3}
For generating a new model for compact stars coupled with isotropic dark energy in the context of  Heintzmann spacetime, we consider the well-known Heintzmann spatial metric potential \cite{hh69} defined as
\begin{eqnarray}
e^{\lambda(r)} &=& \left[1-\frac{3ar^2}{2}\left(\frac{1+C(1+4ar^2)^{-\frac{1}{2}}}{1+ar^2}\right)\right]^{-1}\label{elam}
\end{eqnarray}
where $C$ is dimensionless constant and $a$ is a constant having a dimension of ${\it length^{-2} }$.

Now, from the field equations (\ref{preff})-(\ref{pteff}) we can obtain the following result 
\begin{eqnarray}
  r^2\left(2\nu''(r)+\nu'^2(r)-\nu'(r)\lambda'(r)\right)-2r\left(\nu'(r)+\lambda'(r)\right)+4(e^{\lambda(r)}-1) = 0.  
\end{eqnarray}

After substituting the expression of spatial metric potential (\ref{elam}) in the above equation, we obtain the temporal metric potential $e^{\nu(r)}$ as
\begin{eqnarray}
e^{\nu(r)} &=& A^2(1+ar^2)^3,\label{enu}
\end{eqnarray}

where $A$ is a dimensionless constant.

 Therefore, the metric potentials (\ref{elam}) and (\ref{enu}) produce the solutions of Eqs. (\ref{rhoeff})-(\ref{pteff}) in the following form
\begin{eqnarray}
\rho(r)+\rho^{de}(r) &=& \frac{3 a \left[C \left(3+9 a r^2\right)+\left(3+a r^2\right) \left(1+4 a r^2\right)^{3/2}\right]}{16 \pi  \left(1+a r^2\right)^2 \left(1+4 a r^2\right)^{3/2}},\label{Rheff1}
\\
P(r)+P^{de}(r)&=&\frac{3 a \left[ 3 (1-a r^2) \sqrt{1 + 4 a r^2}-C - 7 a C r^2 \right]}{
 16 \pi(1 + a r^2)^2 \sqrt{1 + 4 a r^2}}.\label{Peff1}
\end{eqnarray}

To find the explicit solutions to the above Eqs. (\ref{Rheff1})-(\ref{Peff1}), we consider the two assumptions followed by the Refs. \cite{cr11, cr05, wb07}

    (i) The dark pressure $P^{de}(r)$ is  proportional to the dark energy density $\rho^{de}(r)$, i.e.,
    \begin{eqnarray}
P^{de}(r) &=& -\rho^{de}(r),\label{rh1}
\end{eqnarray}

(ii)  The dark energy density $\rho^{de}(r)$ is proportional to the normal baryonic matter density $\rho(r)$, i.e.,
\begin{eqnarray}
\rho^{de}(r)   &=& \alpha \rho(r),\label{p1}
\end{eqnarray}

where $\alpha (> 0)$ is a proportionality constant, referred to here as the coupling parameter. It is worth noting that the equation of state for dark energy mentioned above is analogous to the MIT bag model for hadrons \cite{nk12}.

\begin{figure}[!htbp]
\begin{center}
\begin{tabular}{rl}
\includegraphics[width=8.6cm]{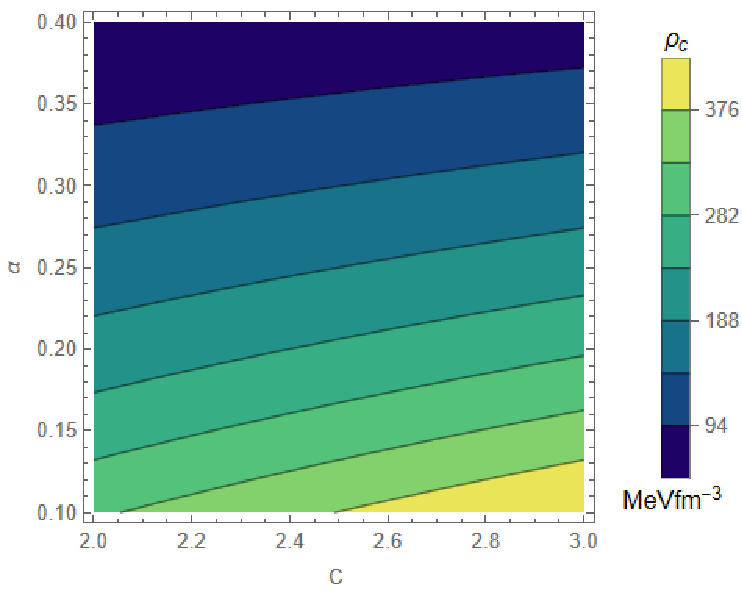}
\includegraphics[width=8.6cm]{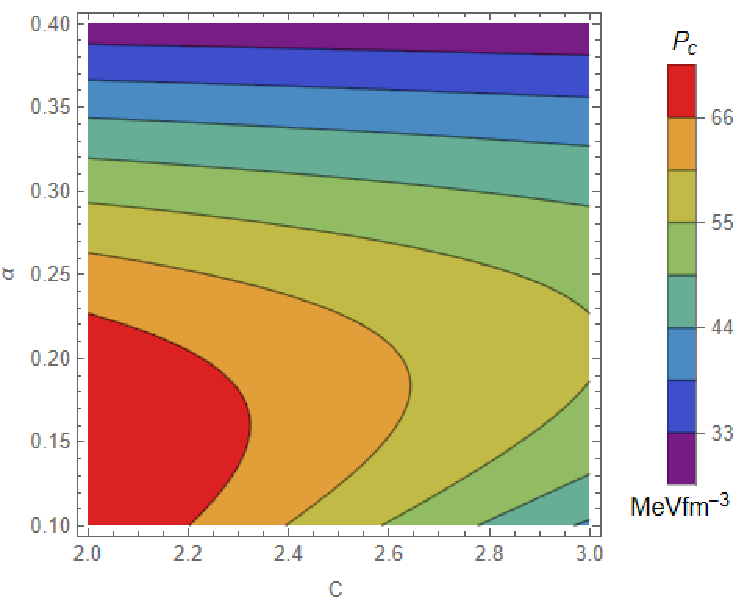}
\\
\end{tabular}
\end{center}
\caption{Contour profiles of central density (Left) and central pressure (Right) against the constants  $C$ and $\alpha$ for the model compact star Vela X-1  associated with the values of constants provided in Table-{\ref{tab1}}.}\label{fig2a}
\end{figure}

Now, on using Eqs. (\ref{rh1})-(\ref{p1}) in Eqs. (\ref{Rheff1})-(\ref{Peff1}), we obtain following results
\begin{eqnarray}
\rho(r) &=&\frac{3 a \left[C \left(3+9 a r^2\right)+\left(3+a r^2\right) \left(1+4 a r^2\right)^{3/2}\right]}{16 \pi  (1+\alpha ) \left(1+a r^2\right)^2 \left(1+4 a r^2\right)^{3/2}} ,\label{rh}
\\
P(r)   &=&\frac{3 a \left[\left(1+4 a r^2\right)^{3/2} \left(3+6 \alpha -ar^2 (3+2 \alpha ) \right)-C \left(1-2 \alpha + 28 a^2 r^4(1+\alpha ) +ar^2 (11+2 \alpha ) \right)\right]}{16 \pi  (1+\alpha) \left(1+a r^2\right)^2 \left(1+4 a r^2\right)^{3/2}} .\label{P}
\end{eqnarray}

Therefore, the energy density and pressure corresponding to the dark energy are obtained as
\begin{eqnarray}
\rho^{de}(r) &=&\frac{3 a\alpha \left[C \left(3+9 a r^2\right)+\left(3+a r^2\right) \left(1+4 a r^2\right)^{3/2}\right]}{16 \pi  (1+\alpha ) \left(1+a r^2\right)^2 \left(1+4 a r^2\right)^{3/2}} ,\label{rhde}
\\
P^{de}(r)&=&-\frac{3 a\alpha \left[C \left(3+9 a r^2\right)+\left(3+a r^2\right) \left(1+4 a r^2\right)^{3/2}\right]}{16 \pi  (1+\alpha ) \left(1+a r^2\right)^2 \left(1+4 a r^2\right)^{3/2}}.\label{Pde}
\end{eqnarray}

Also, the effective energy density, as well as the effective pressure for the current model, are derived as follows
\begin{eqnarray}
\rho^{\text{eff}}(r)=\rho(r)+\rho^{de}(r) &=& \frac{3 a \left[C \left(3+9 a r^2\right)+\left(3+a r^2\right) \left(1+4 a r^2\right)^{3/2}\right]}{16 \pi  \left(1+a r^2\right)^2 \left(1+4 a r^2\right)^{3/2}},\label{reff}
\\
P^{\text{eff}}(r)=P(r)+P^{de}(r)&=&\frac{3 a \left[ 3 (1-a r^2) \sqrt{1 + 4 a r^2}-C - 7 a C r^2 \right]}{
 16 \pi(1 + a r^2)^2 \sqrt{1 + 4 a r^2}}.\label{Peff}
\end{eqnarray}

To know the exact profiles of the model parameters, we will next proceed to determine the values of model constants from the matching conditions of the interior solution with the $Schwarzschild$ exterior solution.

\section{ Determination of constants via matching conditions}\label{Sec4}

Any physical model of a celestial compact object demands the matching of the interior solution with the exterior $Schwarzschild$ solution at the surface $r = R  (> 2M)$  of the compact object. The exterior $Schwarzschild$ solution is given as \cite{ks16}
\begin{eqnarray}
ds^{2} &=& g_{tt}(R)dt^{2}-g_{rr}(R)dr^{2} -r^{2}\big(d\theta^{2}+\sin^{2}\theta d\phi^{2} \big),
\end{eqnarray}
where $g_{tt}(R) = g_{rr}^{-1}(R) = 1-2M/R$. Now, the matching, i.e. the continuity of the metric coefficients between the internal solution and external  {\it Schwarzschild} solution at the boundary $r = R$  yields the following result
\begin{eqnarray}
  A^2 (1 + a R^2)^3 &=& 1-\frac{2M}{R},\label{mc}
 \\
 1-\frac{3 a R^2 \left[1+C(1+4 a R^2)^{-1/2}\right]}{2 \left(1+a R^2\right)} &=& 1-\frac{2M}{R}.\label{mc1}
 \end{eqnarray}


\begin{figure}[!htbp]
\begin{center}
\begin{tabular}{rl}
\includegraphics[width=6cm]{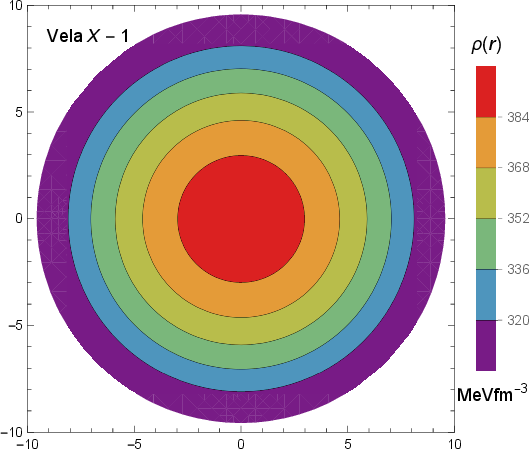}
\includegraphics[width=6cm]{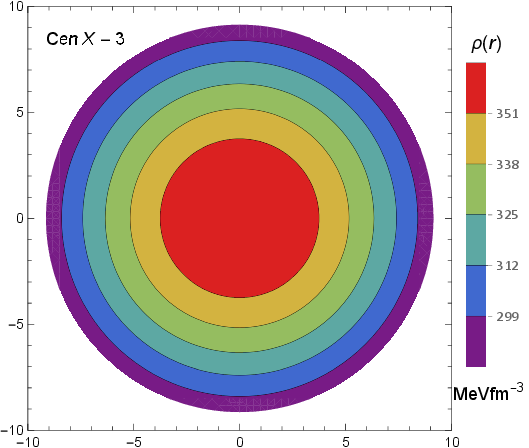}
\includegraphics[width=5.9cm]{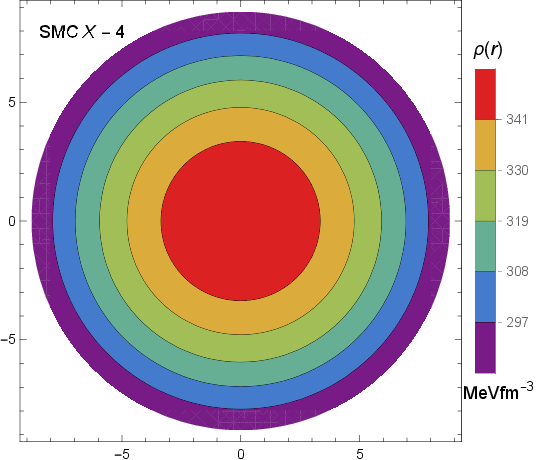}
\\
\end{tabular}
\begin{tabular}{rl}
\includegraphics[width=6.cm]{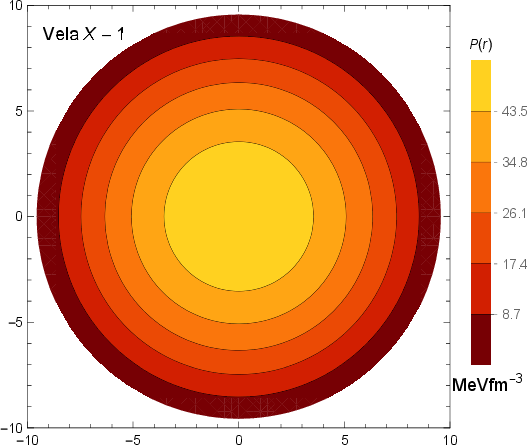}
\includegraphics[width=6cm]{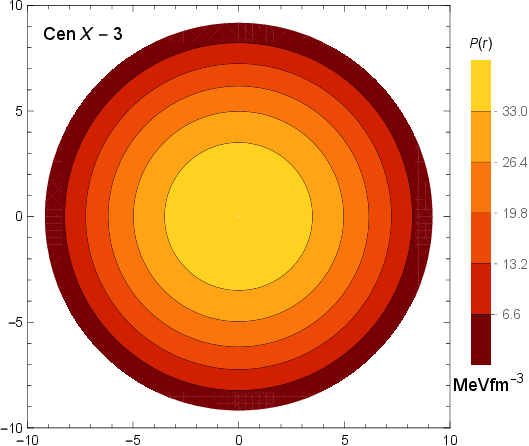}
\includegraphics[width=5.9cm]{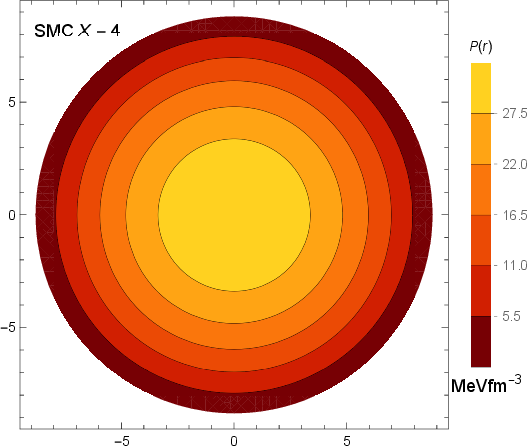}
\\
\end{tabular}
\end{center}
\caption{Radially symmetric profiles of the energy density (Red Centre) and pressure (Yellow Centre)  for the three model compact stars associated with the values of constants provided in Table-{\ref{tab1}}.}\label{fig3}
\end{figure}

 
 Also, the condition that the pressure vanishes at the boundary of the matter configuration, $P(R) = 0$, yields
\begin{eqnarray}
 C\left[1-2 \alpha + 28 a^2R^4 (1+\alpha ) +a R^2 (11+2 \alpha ) \right]+\left(1+4 a R^2\right)^{3/2} \left[a R^2(3+2 \alpha ) -6 \alpha -3\right]=0.\label{mc2}
\end{eqnarray}

Finally, the conditions (\ref{mc})-(\ref{mc2}) provide the expressions  of constants  $a$, $A$ and $C$  in terms of  $\alpha$, radius $R $ and mass $M$  of star in the following form
\begin{eqnarray}
a &=& \frac{M(11 + 2 \alpha )-3R(1+\alpha) +\sqrt{3 M^2 \left(3+52 \alpha +76 \alpha ^2\right) -6 M R\left(3+19 \alpha +22 \alpha ^2\right) +9 R^2(1+\alpha )^2 }}{2 R^2 \left[3 R(4+5 \alpha ) -28 M(1+\alpha ) \right]},
\\
C&=& \frac{\sqrt{1+4 a R^2} \left[4 M \left(1+a R^2\right)-3 a R^3\right]}{3 a R^3},
\\
A&=&\frac{\sqrt{R-2 M}}{\sqrt{R \left(1+a R^2\right)^3}}.
\end{eqnarray}

\begin{figure}[!htbp]
\begin{center}
\begin{tabular}{rl}
\includegraphics[width=8.8cm]{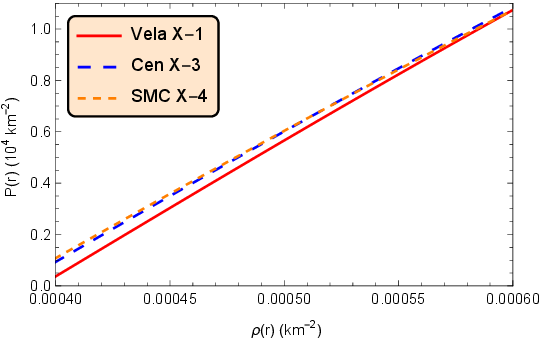}
\includegraphics[width=8.5cm]{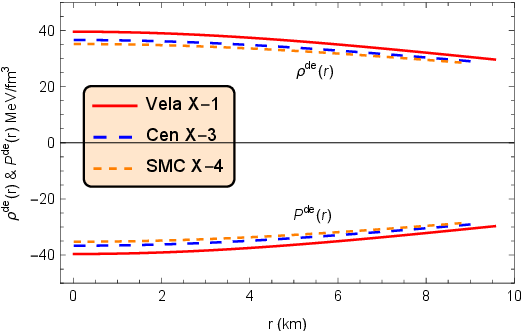}
\\
\end{tabular}
\end{center}
\caption{Profiles of pressure against density (Left) and dark energy density and pressure (Right) against the radial coordinate $r$ for the three model compact stars associated with the values of constants provided in Table-{\ref{tab1}}.}\label{fig4}
\end{figure}

\section{Analysis of physical attributes of  solutions }\label{Sec5}
Here, we aim to demonstrate the physical viability of our model by examining various key conditions in the background of the compact stars Vela X-1, Cen X-3, and SMC X-4.

\subsection{ Profiles of Metric Potentials}

Here, we focus on the behaviours of the temporal metric component $e^{\nu(r)}$ and spatial metric component $e^{\lambda(r)}$. We can easily verify that $e^{\nu(0)} = A^2$, a non-zero constant, and $e^{-\lambda(0)} = 1$, these results ensure that both metric potentials are finite at the centre of the stars. Moreover,  the derivatives of the metric potentials are obtained as
\begin{eqnarray}
  \left(e^{\nu(r)}\right)' &=& \frac{12 a r \left[C-2 a^2 C r^4+2 a C r^2+\left(4 a r^2+1\right)^{3/2}\right]}{\sqrt{4 a r^2+1} \left[a r^2 \left(\sqrt{4 a r^2+1}+3 C\right)-2 \sqrt{4 a r^2+1}\right]^2},
  \\
  \left(e^{\lambda(r)}\right)' &=& 6 a A^2 r \left(1+ar^2\right)^2.
\end{eqnarray}

 One can see that the derivatives of the metric potentials become zero at the centre of the star. The radial profiles of metric potentials, demonstrated in Fig. \ref{fig1} (Left), ensure that they are positive and consistent within the interior of the model stars. Moreover, the reported internal space-time and the Schwarzchild exterior space-time are smoothly matched at the surface $r = R$, i.e. $e^{\nu(R)}$ = $e^{-\lambda(R)}$=$g_{rr}(R)$, which respects the Darmois–Israel condition \cite{cs22, gd27, wi67}. Therefore, the metric potentials are well-behaved within the interior of the stars.

\subsection{ Profiles of Matter Energy Density and Pressure}
In 1984,  Chandrasekhar \cite{sc84} showed that energy density and pressure play a significant role in determining the stability of a compact object against gravitational collapse. Also, the pressure within the compact object can determine the boundary of that compact object \cite{sc84}. In this study, we obtain the central energy density and pressure in the following form
\begin{eqnarray}
\rho_c &=& \rho(0) = \frac{9 a (1+C)}{16 \pi  (1+\alpha )} > 0, \label{rhc}\\
P_c &=& P(0) =  \frac{3 a [3+6 \alpha -C(1-2 \alpha ) ]}{16 \pi  (1+\alpha )} > 0. \label{pc}
\end{eqnarray}

The energy density and pressure are positive and monotonic, decreasing functions of radius $r$ with the maximum value at the centres of the stars, as is clear in Figs. \ref{fig1} (Right) and \ref{fig2} (Left). Also, Fig. \ref{fig2} (Left) ensures that the pressure vanishes at the surfaces of the stars. Here, we obtain the density and pressure gradients in the following form
 \begin{eqnarray}
     \frac{d\rho(r)}{dr} &=& -\frac{3 a^2 r \left[3 C \left(30 a^2 r^4+23 a r^2+5\right)+\left(a r^2+5\right) \left(4 a r^2+1\right)^{5/2}\right]}{8 \pi  (\alpha +1) \left(a r^2+1\right)^3 \left(4 a r^2+1\right)^{5/2}},
     \\
     \frac{dP(r)}{dr} &=& \frac{3 a^2 r \left[\left(4 a r^2+1\right)^{5/2} \left(a (2 \alpha +3) r^2-14 \alpha -9\right)+3 c \xi _1\right]}{8 \pi  (\alpha +1) \left(a r^2+1\right)^3 \left(4 a r^2+1\right)^{5/2}},
 \end{eqnarray}
where,
\begin{eqnarray}
    \xi_1 = 56 a^3 (\alpha +1) r^6-6 a^2 (2 \alpha -3) r^4-a (26 \alpha +3) r^2-6 \alpha -1.
\end{eqnarray}

\begin{figure}[!htbp]
\begin{center}
\begin{tabular}{rl}
\includegraphics[width=8.8cm]{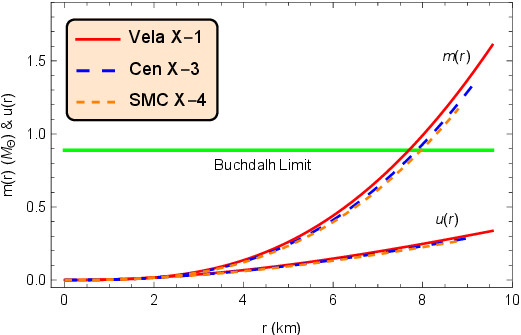}
\includegraphics[width=8.8cm]{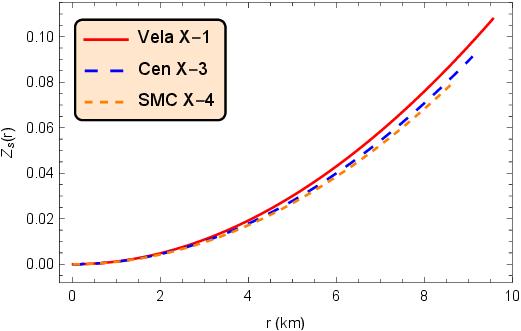}
\\
\end{tabular}
\end{center}
\caption{Profiles of mass and compactness parameter (Left) and surface redshift (Right) against the radial coordinate $r$ for the three model compact stars associated with the values of constants provided in Table-{\ref{tab1}}.}\label{fig5}
\end{figure}

The density and pressure gradients are negative throughout the fluid sphere and vanish at the core, clear in Fig. \ref{fig2} (Right), therefore, these results respect the monotonic decreasing behaviours of the density and pressure. We have also depicted the contour plot for the central energy density and pressure in Fig. \ref{fig2a}, which indicates that central energy density and pressure decrease when the coupling parameter $\alpha$ increases. The symmetric profiles of density and pressure in Fig. \ref{fig3} further confirm that they are well-fitted to their desired properties. Moreover, the pressure of the compact stars is increasing against the density of the stars, Clear in Fig. \ref{fig4} (Left).

\subsection{ Profiles of Dark Energy Density and Dark Pressure}

The dark energy is referred to as "dark" because it lacks an electric charge and does not interact with electromagnetic radiation, including light. Indeed, dark energy is an imaginary type of energy with a very strong negative pressure that tends the accelerating expansion of the universe \cite{pj03}. Recently, dark energy has been used to develop a cyclic model of the universe \cite{lb07}. The present study acknowledges the repulsive nature of dark energy, ensuring that the energy density remains positive while the dark pressure remains negative, as shown in Fig. \ref{fig4} (Right). Moreover, the dark energy density monotonically decreases within the interior of compact objects with the maximum value at the centre, while dark pressure is gradually increasing, clear from Fig. \ref{fig4} (Left). Here, the maximum value of the dark energy is obtained as
\begin{eqnarray}
    \rho^{de}_{\text{max}} = \rho^{de}(0) = \frac{9 a \alpha  (1+C)}{16 \pi  (1+\alpha)} > 0.
\end{eqnarray}

Also, the dark pressure at the centre of the star reduces
\begin{eqnarray}
    P^{de}_{\text{min}} = -\rho^{de}(0) = - \frac{9 a \alpha  (1+C)}{16 \pi  (1+\alpha)} < 0.
\end{eqnarray}

\begin{figure}[!htbp]
\begin{center}
\begin{tabular}{rl}
\includegraphics[width=9cm]{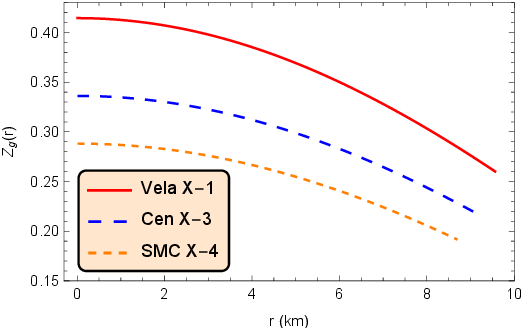}
\includegraphics[width=8.8cm]{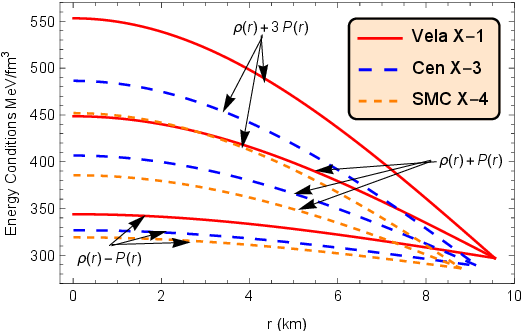}
\\
\end{tabular}
\end{center}
\caption{Profiles of gravitational redshift (Left) and energy conditions (Right) against the radial coordinate $r$ for the three model compact stars associated with the values of constants provided in Table-{\ref{tab1}}.}\label{fig5a}
\end{figure}

\subsection{ Profiles of Mass and Compactness Parameter}
The literature suggests that the active mass of a celestial compact object depends on its density profile, it increases against the confining radius \cite{nk12, ha59}. The mass of a matter configuration  can  be obtained as
\begin{eqnarray}
m(r)&=&4\pi \int_0^r r^{*2} \rho(r^{*}) dr^{*}
=\frac{3 a r^3 \left(1+4 a r^2+C \sqrt{4 a r^2+1}\right)}{4 (1+\alpha ) \left(1+5 a r^2+4 a^2 r^4\right)} ,
\end{eqnarray}

The above result of the mass function $m(r)$ ensures that  $m(0) = 0$. We have demonstrated the mass function $m(r)$ against the radius $r$ in Fig. \ref{fig5} (Left), which indicates that the mass function is regular, monotonic increasing within the interior of the stars and the maximum mass is attained at the boundary $r = R$.

Also, the compactness parameter is expressed in terms of the mass function as
\begin{eqnarray}
u(r)&=&\frac{2m(r)}{r} = \frac{3 a r^2 \left(1+4 a r^2+C \sqrt{4 a r^2+1}\right)}{2 (1+\alpha ) \left(1+5 a r^2+4 a^2 r^4\right)}
\end{eqnarray}.

The profile of the compactness parameter is displayed in Fig. \ref{fig5} (Left). It gradually increases with the radius $r$ and respects the Buchdahl Limit, i.e. $u(r) < 8/9$ \cite{ha59}.

\subsection{ Profiles of  Surface Redshift and Gravitational Redshift}

The surface redshift is determined by the stellar mass and radius, or in other words, by the surface gravity.
For our model, the surface redshift $Z_s(r)$  is obtained as
\begin{eqnarray}
Z_s(r)&=&\left[1-u(r)\right]^{-1/2}-1=\left[1-\frac{3 a r^2 \left(1+4 a r^2+C \sqrt{1+4 a r^2}\right)}{2 (1+\alpha ) \left(1+5 a r^2+4 a^2 r^4\right)}\right]^{-1/2}-1.
\end{eqnarray}

 We depict the profile of $ Z_S (r)$ in Fig. \ref{fig5} (Right) from the centre to the surface of the stars. The surface redshift is gradually increasing with $r$ by attaining the maximum at the surface. 
 
Furthermore, the gravitational redshift $Z_g(r)$ is obtained as
\begin{eqnarray}
Z_g(r)&=&e^{-\nu(r)/2}-1 =\frac{1}{A}\left(a r^2+1\right)^{-3/2}-1 .
\end{eqnarray}

The gravitational redshift $Z_g(r)$ has the opposite trend of $ Z_S (r)$, it is maximum at the centre and gradually decreases towards the surface by reaching its minimum at the surface, clear in Fig. \ref{fig5a} (Left). It is important to note that neither type of redshift exhibits a singularity throughout the matter configurations.

\begin{figure}[!htbp]
\begin{center}
\begin{tabular}{rl}
\includegraphics[width=8.8cm]{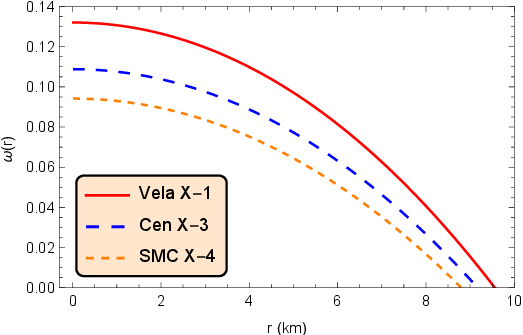}
\includegraphics[width=8.6cm]{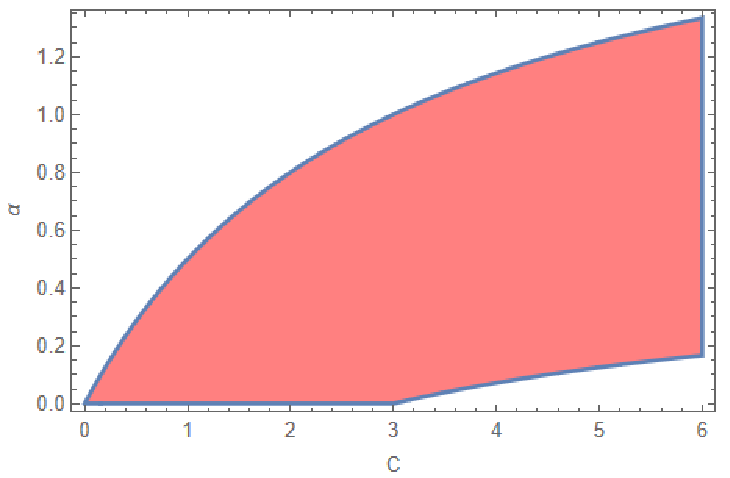}
\\
\end{tabular}
\end{center}
\caption{Profiles of EoS parameter $\omega(r)$ against the radial coordinate $r$ for the three model compact stars associated with the values of constants provided in Table-{\ref{tab1}} (Left)  and constraint region of coupling parameter $\alpha$ against $C$ (Right).}\label{fig6}
\end{figure}

\subsection{ Profiles of Energy Conditions } 

For the physical stellar model, it is necessary to examine whether the present solutions satisfy all the energy conditions inside the compact objects or not. There are four primary types of energy conditions, namely  \textbf{(i)} Null energy condition (NEC), \textbf{(ii)} Weak energy condition (WEC), \textbf{(iii)} Dominant energy condition (DEC), and \textbf{(iv)} Strong energy condition (SEC), these are defined as \cite{step, wald}
\begin{eqnarray}
\text{NEC} &:&  \rho(r)+P(r) \geq  0, ~~~~~~\text{WEC} : \rho(r) \geq  0,~\rho(r)+P(r) \ge 0,
\\
\text{DEC} &:& \rho(r) \geq  0,~\rho(r)-P(r) \ge 0,~~~~\text{SEC} :  \rho(r)+3P(r) \ge 0.\label{EC}
\end{eqnarray}

The present solutions yield the following results 
\begin{eqnarray}
\rho(r)+P(r) &=&\frac{3 a \left[C \left(1-14 a^2 r^4-a r^2\right)+\left(3-4 a^2 r^4+11 a r^2\right) \sqrt{4 a r^2+1}\right]}{8 \pi  \left(1+a r^2\right)^2 \left(1+4 a r^2\right)^{3/2}} ,
\\
\rho(r)-P(r) &=& \frac{3 a \left[C \left(14 a^2 (\alpha +1) r^4+a (\alpha +10) r^2-\alpha +2\right)+\left(4 a r^2+1\right)^{3/2} \left(a (\alpha +2) r^2-3 \alpha \right)\right]}{8 \pi  (\alpha +1) \left(a r^2+1\right)^2 \left(4 a r^2+1\right)^{3/2}},
\end{eqnarray}
\begin{eqnarray}
\rho(r)+3P(r) &=& \frac{3a}{8 \pi  (1+\alpha ) \left(1+a r^2\right)^2 \left(1+4 a r^2\right)^{3/2}}\bigg[3 \left((2+3 \alpha ) \sqrt{1+4 a r^2}+\alpha  C\right)-2 a^2 r^4 \Big(2 (4+3 \alpha ) \sqrt{1+4 a r^2}\nonumber
\\
&&+21 (1+\alpha ) C\Big)-a r^2 \left(3 (4+\alpha ) C-(33 \alpha +20) \sqrt{1+4 a r^2}\right)\bigg].
\end{eqnarray}

The graphical representations in  Fig. \ref{fig1} (Right) and Fig. \ref{fig5a} (Right) ensure that the present solutions satisfy all the energy conditions inside the matter configurations. Therefore, the present model is physically realistic in nature.

\begin{figure}[!htbp]
\begin{center}
\begin{tabular}{rl}
\includegraphics[width=8.8cm]{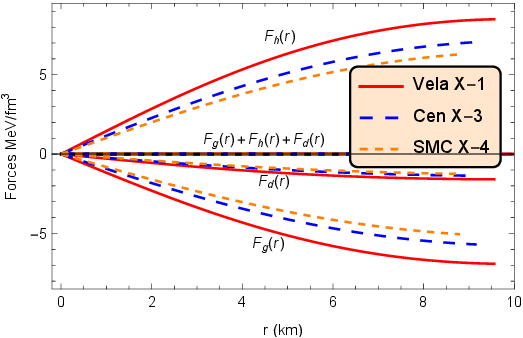}
\includegraphics[width=8.8cm]{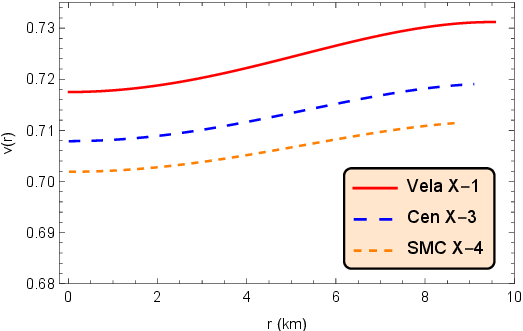}
\end{tabular}
\end{center}
\caption{Profiles of different forces (Left) and sound velocity (Right) against the radial coordinate $r$ for the three model compact stars associated with the values of constants provided in Table-{\ref{tab1}}.}\label{fig7}
\end{figure}

\subsection{ Profiles of Equation of State via Zel’dovich Condition}  

The physical stellar structure must satisfy the Zel’dovich requirement \cite{sl08, yb71, yb72, yl62}, which states that the ratio of pressure to density must be less than unity throughout the stellar interior, mathematically, $\omega(r) = P(r)/\rho(r) < 1$ for $r < R$. Interestingly, our reported solutions respect the Zel’dovich requirement, clear from Fig. \ref{fig6} (Left) and hence the solutions representing matter distributions are physical.

Now, the Zel’dovich condition on the central values of pressure and density yields $P_{c}/ \rho_c < 1$, from which we obtain the following relationship
\begin{eqnarray}
\alpha < \frac{2C}{3+C}. \label{zel}
\end{eqnarray}

Also, the relation $P_c > 0$ yields
\begin{eqnarray}
\alpha > \frac{C-3}{2(C+3)}. \label{pc}
\end{eqnarray}

Therefore, the above results (\ref{zel})-(\ref{pc}) generates a boundary representing constraint on the  coupling parameter $\alpha$ as
\begin{eqnarray}
\frac{C-3}{2(C+3)} < \alpha < \frac{2C}{3+C}.\label{cont}
\end{eqnarray}

In Fig. \ref{fig6} (Right), we display the constraint region for the coupling parameter $\alpha$ given in the expression (\ref{cont}) with respect to the constant $C$.

\section{EQUILIBRIUM analysis}\label{Sec6}
Here, we examine the equilibrium situation of the model stars with the help of the generalised {\it Tolman-Oppenheimer-Volkoff} (TOV) equation \cite{jr39}. Indeed, the equilibrium situation occurs due to the dynamic balancing of interior forces. For our model,  the generalised TOV equation can be written as \cite{PB21}
\begin{eqnarray}
&&-\frac{\nu'(r)}{2}[\rho(r)+P(r)]-\frac{dP(r)}{dr}-\frac{dP^{de}(r)}{dr}=0.  \label{to5}
\end{eqnarray}

The above equation  can be written as
\begin{eqnarray}
F_g(r)+F_h(r)+F_d(r)=0,  \label{to6}
\end{eqnarray}
where,
\begin{eqnarray}
F_g(r) &=& -\frac{\nu'(r)}{2}[\rho(r)+P(r)], ~\text{termed as the gravitation force },\nonumber
\\
F_h(r) &=& -\frac{dP(r)}{dr},~\text{termed as the hydrostratics force },\nonumber
\\
F_d(r) &=& -\frac{dP^{de}(r)}{dr}, ~\text{termed as the force due to dark energy}.
\end{eqnarray}

\begin{figure}[!htbp]
\begin{center}
\begin{tabular}{rl}
\includegraphics[width=8.8cm]{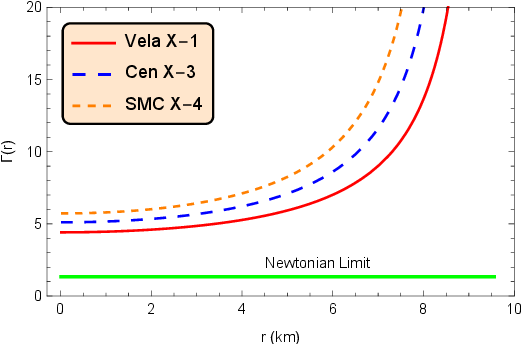}
\includegraphics[width=8.8cm]{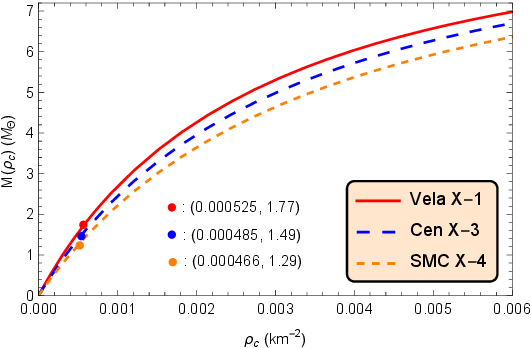}
\end{tabular}
\end{center}
\caption{Profiles of adiabatic index against the radial coordinate $r$ (Left) and mass against the central density $\rho_c$ for the three model compact stars associated with the values of constants provided in Table-{\ref{tab1}}.}\label{fig8}
\end{figure}

For the solutions presented in this study, the expressions for the three forces are obtained as follows 
\begin{eqnarray}
 F_g(r) &=& \frac{9 a^2 r \left[C \left(14 a^2 r^4+a r^2-1\right)+\left(4 a^2 r^4-11 a r^2-3\right) \sqrt{1+4 a r^2}\right]}{8 \pi  \left(1+a r^2\right)^3 \left(1+4 a r^2\right)^{3/2}},
\\
 F_h(r) &=& \frac{3 a^2 r }{8 \pi  (1+\alpha ) \left(1+a r^2\right)^3 \left(1+4 a r^2\right)^{5/2}}\bigg[3 C \left(-56 a^3r^6 (1+\alpha ) +6 a^2r^4 (2 \alpha -3) +a r^2(3+26 \alpha ) +6 \alpha +1\right)\nonumber
 \\
 &&+\left(1+4 a r^2\right)^{5/2} \left(9+14 \alpha -ar^2 (3+2 \alpha ) \right)\bigg],
\\
 F_d(r) &=&-\frac{3 a^2 \alpha  r \left[3 C \left(5+23 a r^2+30 a^2 r^4\right)+\left(5+a r^2\right) \left(1+4 a r^2\right)^{5/2}\right]}{8 \pi  (1+\alpha ) \left(1+a r^2\right)^3 \left(1+4 a r^2\right)^{5/2}} .
\end{eqnarray}

The profiles of these three forces are graphically demonstrated in Fig. \ref{fig7} (Left), which indicates that the present solutions respect the TOV equation, i.e. $F_g(r)+F_h(r)+F_d(r)$ remains zero throughout the interior of the model stars. Therefore, the model stars are maintaining their equilibrium position in the background of our reported solutions. 

\section{ STABILITY ANALYSIS}\label{Sec7}
In this section, we analyse the stability of the present model via sound velocity, adiabatic index, and {\it Harrison-Zeldovich-Novikov's} static stability condition.

\subsection{ Sound Velocity}
 
At the time of propagation of sound through a physical stellar matter configuration, its velocity must be less than the velocity of light $c$, as the velocity of light is the maximum velocity ever.  Mathematically, sound velocity $V(r)$: $0< V(r) < c$ (= 1), this condition is known as the causality condition. It is important to note that a matter sphere that does not satisfy the causality condition is considered nonphysical. The sound velocity $V(r)$ is defined as
\begin{eqnarray}
V(r) &=& \left[{dP(r) \over d\rho(r)}\right]^{1/2} =\left[\frac{3 C \left(1+6 \alpha +6 a^2 r^4(2 \alpha -3) -56 a^3r^6 (1+\alpha ) +a  r^2(3+26 \alpha )\right)+\xi_2}{3 C \left(30 a^2 r^4+23 a r^2+5\right)+\left(5+a r^2\right) \left(1+4 a r^2\right)^{\frac{5}{2}}}\right]^{\frac{1}{2}},
\end{eqnarray}

where,
\begin{eqnarray}
    \xi_2 = \left(1+4 a r^2\right)^{\frac{5}{2}} \left(9+14 \alpha-ar^2 (2 \alpha +3) \right).
\end{eqnarray}

Fig. \ref{fig7} (Right) ensures that the reported solutions satisfy the causality condition, and hence, the representing matter distributions are physical.

\begin{figure}[!htbp]
\begin{center}
\begin{tabular}{rl}
\includegraphics[width=8.8cm]{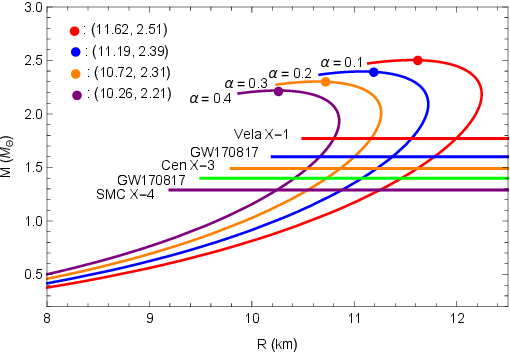}
\includegraphics[width=8.8cm]{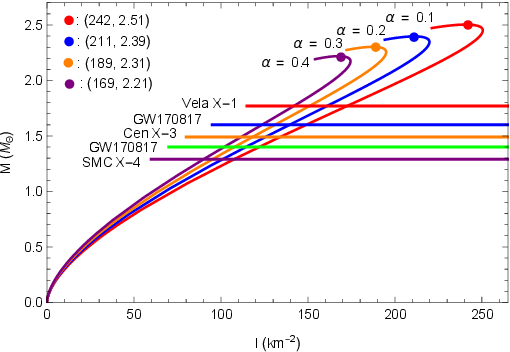}
\end{tabular}
\end{center}
\caption{Profiles of mass against the surface radius $R$ (Left) and mass against the moment of inertia $I$.}\label{fig9}
\end{figure}

\subsection{Adiabatic Index}

The thermal stability of the matter sphere against gravitational collapse depends on the adiabatic index $\Gamma(r)$. The literature ensures that the thermal stability demands $\Gamma(r) > 4/3$ (Newtonian limit) within the interior of the matter sphere \cite{hb64, hh75, rc93, chc17}. The adiabatic index $\Gamma(r)$ can be expressed in terms of the energy density, pressure, and sound velocity as follows
\begin{eqnarray}
\Gamma(r)&=& \left(1+\frac{\rho(r)}{P(r)}\right)V^2(r)=\frac{\bigg[3 C \left(a r^2 \left(3-2 a r^4 \left(9+28 a r^2(\alpha +1) -6 \alpha \right)+26 \alpha \right)+6 \alpha +1\right)+\xi_2\bigg]\xi_3}{3 C \left(5+a r^2 (30 a r^2+23)\right)+\left(5+a r^2\right) \left(1+4 a r^2\right)^{5/2}},
\end{eqnarray}
where,
\begin{eqnarray}
\xi_3=1-\frac{C \left(3+9 a r^2\right)+\left(3+a r^2\right) \left(1+4 a r^2\right)^{3/2}}{C \left(a r^2 \left(28 ar^2 (\alpha +1) +2 \alpha +11\right)-2 \alpha +1\right)+\left(1+4 a r^2\right)^{3/2} \left(ar^2 (2 \alpha +3) -6 \alpha -3\right)}.
\end{eqnarray}

The profile $\Gamma(r)$ is depicted in Fig. \ref{fig8} (Left). Interestingly,  the adiabatic index nicely satisfies its desired range to hold the stable matter spheres against gravitational collapse.

\subsection{Harrison-Zeldovich-Novikov's Static Stability Condition}

As previously mentioned, energy density plays a crucial role in the stability of the matter sphere. In this regard, Harrison et al. \cite{har65} and Zeldovich-Novikov \cite{yb72} proposed that the condition for static stability as $\partial M(\rho_c)/\partial \rho_c > 0$, meaning that the mass of the matter sphere should increase with the central density. Otherwise, the matter sphere becomes unstable. For our model, we obtain the mass in terms of central density $\rho_c$ in the following form
\begin{eqnarray}
 M(\rho_c)&=& \frac{\pi  \rho_c R^3 \left[C \sqrt{82944 \pi \rho_c R^2 (1+\alpha ) (1+C) +11664 (1+C)^2}+108 (1+C)+768 \pi \rho_c R^2 (1+\alpha ) \right]}{16 \pi\rho_c R^2  (1+\alpha )  \left(45 C+64 \pi\rho_c R^2  (1+\alpha ) +45\right)+81 (1+C)^2}.
 \end{eqnarray}

The profile of $M(\rho_c)$ in Fig. \ref{fig8} (Right) clearly shows that it is increasing against the central pressure of the model stars, and hence, $\partial M(\rho_c)/\partial \rho_c > 0$. Therefore, the model stars are stable under the static stability condition.

\begin{table*}[thth]
\caption{ Numerical values of masses, radii, and model constants for three compact stars with their average density corresponding to $\alpha = 0.1$.}
\label{tab1}       
\centering
\begin{tabular}{|c|c|c|c|c|c|c|c|c|c|c}
\hline
Star & $M\slash M_{\odot}$ &   $R ~(km)$   &  Ref.   &  $M\slash M_{\odot}$ &   $R ~(km)$ & $a ~(km^{-2})$ & $C$  & $A$ &  $\rho_{av}=3M_{cal}/4\pi R^3_{cal} $  \\
    & Observed           &   Observed         &         &     Calculated         & Calculated  & & & &  $(gm/cm^{3})$. \\
    \hline
Vela X-1     & 1.77 $\pm$ 0.08 &  9.560 $\pm$ 0.08 &  \cite{ml11}   & 1.77   & 9.56 & 0.000876 & 2.6783 & 0.706982 &  6.51464 $\times 10^{14}$  \\  
    \hline
Cen X-3      & 1.49 $\pm$ 0.08 &  9.170 $\pm$ 0.13 &  \cite{ml11}   &  1.49  & 9.17 & 0.000763 & 2.90605 & 0.748407 &  6.21397 $\times 10^{14}$   \\
    \hline
SMC X-4 & 1.29 $\pm$ 0.05 &  8.831 $\pm$ 0.09 &  \cite{ml11}   &  1.29  & 8.80 & 0.000705 & 3.06458 &  0.776308 &  6.08741 $\times 10^{14}$   \\
\hline
\end{tabular}\label{table3}
\end{table*}

\begin{table*}[thth]
\caption{Numerical values of central density, central pressure, surface density, surface gravitational redshift, and surface compactness parameter for the compact stars given in Table-\ref{tab1}.}\label{tab2}

\centering
\begin{tabular}{|c|c|c|c|c|c|c|c|cc}
\hline
Star   &  $\rho_c ~ (gm/cm^3)$ & $\rho(R) ~(gm/cm^3)$ & $P_c ~(gm/cm^2)$ &  $z(R)$ & $u(R)$ & Buchdahl Limit\cite{ha59}\\
\hline
Vela X-1    &    7.06333 $\times 10^{14}$ & 5.29476 $\times 10^{14}$ & 8.38387 $\times 10^{34}$ & 0.22778 & 0.33663 & $< 8/9$\\
\hline
Cen X-3  &   6.53669 $\times 10^{14}$ & 5.14436 $\times 10^{14}$ & 6.39289 $\times 10^{34}$& 0.19135 & 0.29543 & $< 8/9$\\
\hline
SMC X-4   &  6.28134 $\times 10^{14}$& 5.10003 $\times 10^{14}$ & 5.31644 $\times 10^{34}$& 0.16764 & 0.26653 & $< 8/9$\\
\hline
\end{tabular}
\end{table*}


\begin{table*}[thth]
\caption{ Numerical values of predicted radii $R$ for five well-known compact stars corresponding to $\alpha$ =\{0.1, 0.2, 0.3, 0.4\}.}
\label{tab3}       
\centering
\begin{tabular}{| *{7}{c|}}
\hline
  Star  & Observed &  Refs.    
                    & \multicolumn{4}{c|}{Predicted $R$ (km)} 
                            \\
                            \cline{4-7}
  &  $M~ (M_{\odot})$   &       &      $\alpha = 0.1$  & $\alpha= 0.2$ & $\alpha = 0.3$ &  $\alpha = 0.4$ \\
\hline
Vela X-1     & 1.77 &    \cite{ml11}  &  11.99 & 11.58  & 11.18 & 10.79\\  
\hline
GW170817     &  1.6 &   \cite{ab17} & 11.79 & 11.41  & 11.01 & 10.67\\
\hline
Cen X-3     & 1.49 &    \cite{ml11}    & 11.64& 11.23 & 10.87 & 10.57\\
\hline
GW170817 & 1.4 &   \cite{cd20}   & 11.48 & 11.12 & 10.76 & 10.43\\
\hline
SMC X-4      & 1.29 & \cite{ml11} & 11.26 & 10.88  & 10.55  & 10.26\\
\hline
\end{tabular}\label{table3}
\end{table*}

\section{Maximum mass-radii and Moment of inertia }\label{Sec8}

This section analyses our solutions representing maximum mass to incorporate with the recent observational data. In this context, we have plotted the $M-R$ relationship in Fig. \ref{fig9} (Left) and fitted the observational data for five compact stars to predict their radii from the $M-R$ graph. Here, it is observed that the maximum mass in the $M-R$ graph decreases as the coupling parameter $\alpha$ increases. The present solutions represent the maximum mass $M_{max} = $ 2.51 $M_\odot$, 2.39 $M_\odot$, 2.31 $M_\odot$, and 2.21 $M_\odot$ with the surface radius $R_s =$ 11.62 $km$, 11.19 $km$, 10.72 $km$, and 10.26 $km$ corresponding to $\alpha =$ 0.1, 0.2, 0.3, and 0.4, respectively. It is worth noting that all the maximum masses are under the  Rhoades-Ruffini limit $3.2 M_{\odot}$ \cite{ruff}. The predicted radii for the five compact stars are given in Table-\ref{tab3} corresponding to $\alpha =$ 0.1, 0.2, 0.3, and 0.4. The observational results from GW170817 claimed that the neutron stars of masses 1.6 $M_{\odot}$ and 1.4 $M_{\odot}$ must have radius at least $10.68^{+0.15}_ {-0.04}~ km$ \cite{ab17} and $11^{+0.9}_ {-0.6}~ km$ \cite{cd20}. Interestingly, Table-\ref{tab3} shows that a star with a mass of 1.6 $M_{\odot}$ fits well for $\alpha$ = 0.1, 0.2, and 0.3, while a star with a mass of 1.4 $M_{\odot}$ fits well for $\alpha$ = 0.1 and $\alpha$ = 0.2.

The moment of inertia for a uniformly rotating matter sphere  can be expressed as \cite{latt}
\begin{eqnarray}
I = {8\pi \over 3} \int_0^R r^4 \left[\rho(r)+P(r)\right] e^{[\lambda(r)-\nu(r)]/2} ~{\bar{\omega} \over \Omega}~dr.
\end{eqnarray}

Here, $\Omega$ represents the angular velocity of the matter sphere  and $\bar{\omega}$ represents the rotational drag satisfying Hartle's equation, given by \cite{hart} 
\begin{eqnarray}
{d \over dr} \left(r^4 \mathcal{H} ~{d\bar{\omega} \over dr} \right) =-4r^3\bar{\omega}~ {d\mathcal{H} \over dr} .
\end{eqnarray}
where $\mathcal{H} =e^{-[\lambda(r)+\nu(r)]/2}$ with $\mathcal{H} (R)=1$. In this regard, Bejger and Haensel \cite{bejg} proposed an approximate solution for the moment of inertia $I$ up to the maximum mass $M_{max}$, given as
\begin{equation}
I = {2 \over 5} \Big(1+x\Big) {MR^2},
\end{equation}
with the parameter $x = (M/R)\cdot km/M_\odot$.  The precise behaviour of the moment of inertia $I$ as a function of mass is illustrated in Fig. \ref{fig9} (Right). It reveals that the moment of inertia increases with mass up to a certain point, after which it begins to decrease. The dot sign in the $I-M$ graph indicates that the values of $I$ at the $M_{max}$ are 242 $km^{-2}$, 211 $km^{-2}$, 189 $km^{-2}$, and 169 $km^{-2}$ for $\alpha =$ 0.1, 0.2, 0.3, and 0.4, respectively. It is important to note here that  $I_{max}$ are not attending at $M_{max}$, however,   $I_{max}$ = 250 $km^{-2}$, 219 $km^{-2}$, 194 $km^{-2}$, and 174 $km^{-2}$ at $M $ = 2.43 $M_\odot$, 2.32 $M_\odot$, 2.22 $M_\odot$, and 2.15 $M_\odot$ for $\alpha =$ 0.1, 0.2, 0.3, and 0.4, respectively.

\section{Results and Conclusion}\label{Sec9}
In this article, we have attempted to develop a distinctive model for celestial compact stars coupled with isotropic dark energy in Heintzmann spacetime in the context of Einstein's gravity. In order to solve the Einstein field equations for a static spherically symmetric perfect matter configuration, we have considered the well-known Heintzmann ansatz and a particular form of EoS that relates the baryonic matter with the dark energy depending on the coupling parameter $\alpha$. As mentioned above, numerous studies have previously been conducted on celestial compact stars composed of ordinary fluid and dark energy. In this theoretical investigation, we have focused on the study of structural parameters, equilibrium, stability, and  $M–R$ relations for the choices of well-known compact stars Vela X-1, Cen X-3, and SMC X-4, along with the numerical results for structural parameters. For a comprehensive analysis, we have matched our interior solution with the exterior Schwarzschild solution, which helps to determine the values of the model constants. We have also compared our numerical results with available observational data to ensure the physical validity of the model. The key features of our proposed dark energy stellar model are as follows:


 {\bf1.} The metric potential functions $e^{\nu(r)}$ and $e^{-\lambda(r)}$ are finite and regular over the radius of the stars, clear from \ref{fig1} (Left). Moreover, the internal metric potentials meet with the external Schwarzschild solution at the surface of stars. Thus, all these results confirm that the Heintzmann metric potential is appropriate for constructing a regular compact stars model coupled with isotropic dark energy. 
 
 {\bf2.} The energy density $\rho(r)$ and pressure $P(r)$ both are positive, finite, and decreasing towards the surface of the stars, clear from  Fig. \ref{fig1} (Right) and Fig. \ref{fig2} (Left). The density and pressure reach their maximum levels at the core, ensuring the presence of highly compact cores. Moreover, the pressure becomes zero at the surface i.e. $P(R)$ = 0, and the profiles of density and pressure gradients also confirmed the decreasing nature of density and pressure within the model stars, Fig. \ref{fig2} (Right). We have also analyzed the central density and central pressure for the star Vela X-1 against the coupling constant $\alpha$ in Fig. \ref{fig2a}, which indicates that both central density and central pressure reduce for increasing values of $\alpha$. Hence, the coupling constant $\alpha$ exhibits an inverse relationship with both the density and pressure. The symmetric density and pressure profiles in Fig. \ref{fig3} further confirm that they are well-fitted to display their desired properties. The numerical values of central and surface densities are on the order of $10^{14}~gm/cm^3$, and central pressure is on the order of $10^{34}~gm/cm^2$ for the model stars, given in  Table-\ref{tab2}, match well with the observational data. In addition, we have depicted the profile of pressure and the energy density in Fig. \ref{fig4} (Left), indicating increasing characteristics of pressure with increased energy density. In this model, the dark energy density profile is positive and decreases with the radius of stars, while dark pressure has the opposite behaviour, hence, this model represents the dark energy with its heredity property of repulsive character, clear from Fig. \ref{fig4} (Right). 

{\bf 3.}  The mass function and compactness parameter are both positively finite, regular, and exhibit an increasing trend r throughout the interior of stars, clear from Fig. \ref{fig5} (Left). Moreover, the compactness parameter is well-fitted under the Buchdahl limit for each of the model stars, clear from Fig. \ref{fig5} (Left)  and Table-\ref{tab2}. In this model, the surface redshift $Z_s(r)$  is minimum at the centre and then gradually increasing towards the surface, clear from Fig. \ref{fig5} (Right), in contrast, the gravitational redshift $Z_g(r)$ exhibiting the behaviours in the opposite of $Z_s(r)$ (See Fig. \ref{fig5a} (Left)). The numerical results of the $Z_s(r)$ at the surface of model stars are provided in Table-\ref{tab2}, interestingly, all these values are under the proper range \cite{bv02}.

{\bf 5.} The literature suggests that a compact star composed of physical matter must fulfil the NEC, WEC, and SEC throughout its interior. The present solutions have satisfied all the energy conditions within the interior of model stars, as is clear in Figs. \ref{fig1} (Left) and \ref{fig5a} (Right). Therefore, the solutions describing matter spheres are physical in behaviour. Moreover, the physical behaviour of matter spheres is also confirmed by the range of EoS parameter $0 < \omega (r) <1$, clear from Fig. \ref{fig6} (left). 

{\bf 6.} The reported solutions have satisfied the generalised TOV equations within the interior of the model stars, clear from Fig. \ref{fig7} (Left). Thus, our solutions are well-fitted for the model stars to maintain their equilibrium position under the simultaneous action of gravitational, hydrostatic, and force due to dark energy.

{\bf 7.} The solutions once again support a physical matter distribution by adhering to the causality conditions (See Fig. \ref{fig7} (Right)). Besides,  the adiabatic index $\Gamma_r(r)>$ $\frac{4}{3}$ (See Fig. \ref{fig8} (Left)) and the mass profile increases for increasing central density  (See Fig. \ref{fig8} (Right)). Therefore, all these results ensure that our solutions nicely support the static stable physical matter configurations.  Fig. \ref{fig8} (Right) also shows that the central density for model stars Vela X-1, Cen  X-3, and SMC-X-4 are 0.000525 $km^{-2}$, 0.000485 $km^{-2}$, and 0.000466 $km^{-2}$, respectively.

{\bf 8.} The solutions producing maximum masses $M$ in relation to the surface radius $R$ are illustrated in Fig. \ref{fig9} (Left) for coupling parameter $\alpha$ = 0.1, 0.2, 0.3, 0.4, demonstrating that the maximum mass decreases as $\alpha$ increases. The current solutions yield maximum masses of $M_{max} = $ 2.51 $M_\odot$, 2.39 $M_\odot$, 2.31 $M_\odot$, and 2.21 $M_\odot$ with surface radii of $R_s =$ 11.62 $km$, 11.19 $km$, 10.72 $km$, and 10.26 $km$ for $\alpha =$ 0.1, 0.2, 0.3, and 0.4, respectively. It is worth noting that all maximum masses remain below the Rhoades-Ruffini limit of $3.2 M_{\odot}$ \cite{ruff}. Moreover, we have estimated the predicted radii for the five compact stars for $\alpha =$ 0.1, 0.2, 0.3, and 0.4 from Fig. \ref{fig9} (Left), given in Table-\ref{tab3}. The observational results from GW170817 suggest that neutron stars with masses of 1.6 $M_{\odot}$ and 1.4 $M_{\odot}$ must have radii of at least $10.68^{+0.15}_{-0.04}$ km \cite{ab17} and $11^{+0.9}_{-0.6}$ km \cite{cd20}, respectively. In our model, one can easily check from Table-\ref{tab3} a star with a mass of 1.6 $M_{\odot}$ fits well for $\alpha = 0.1$, $\alpha = 0.2$, and $\alpha = 0.3$, while a star with a mass of 1.4 $M_{\odot}$ fits well for $\alpha = 0.1$ and $\alpha = 0.2$.

{\bf 9.} As the mass $M$ increases, the moment of inertia $I$ initially rises, reaching a peak at a certain mass value before it begins to decrease (See Fig. \ref{fig9} (Right)). Notably, the mass is slightly lower at $I_{max}$ compared to the maximum mass, $M_{max}$. In this study, the numerical values of $I$ at the $M_{max}$ are 242 $km^{-2}$, 211 $km^{-2}$, 189 $km^{-2}$, and 169 $km^{-2}$ for $\alpha =$ 0.1, 0.2, 0.3, and 0.4, respectively. It is important to note here that  $I_{max}$ are not attending at $M_{max}$, however,   $I_{max}$ = 250 $km^{-2}$, 219 $km^{-2}$, 194 $km^{-2}$, and 174 $km^{-2}$ at $M $ = 2.43 $M_\odot$, 2.32 $M_\odot$, 2.22 $M_\odot$, and 2.15 $M_\odot$ for $\alpha =$ 0.1, 0.2, 0.3, and 0.4, respectively.


 Finally, all the significant findings confirm that the present study meets all the necessary conditions to present a physically acceptable, stable, and singularity-free generalised model for compact stars coupled with dark energy in the framework of Einstein's gravity.  We anticipate that our model could hypothetically offer valuable insights into astrophysical phenomena on a larger scale.


\begin{thebibliography}{99}


\bibitem{ks16} K. Schwarzschild,  {\it Sitzber. Deut. Akad. Wiss. Berlin, Kl. Math.Phys. Tech.}, 424-434 (1916).
\bibitem{rh26} R. H. Fowler, {\it Mon. Not. R. Astron. Soc.} {\bf 87}, 114 (1926).
\bibitem{rc39} R. C. Tolman,  {\it Phys. Rev.} {\bf 55}, 364 (1939).
\bibitem{jr39} J. R. Oppenheimer and G. Volkoff, {\it Phys. Rev.} {\bf 55}, 374 (1939).
\bibitem{sc31} S. Chandrasekhar, {\it Astrophys. J.} {\bf 74}, 81 (1931).
\bibitem{bp17} B. P. Abbott et al., {\it Physical Review Letters} {\bf 119}, 161101, (2017).
\bibitem{ra20} R. Abbott et al., {\it Astrophysical Journal Letters} {\bf 896}, 44 (2020).
\bibitem{zy18} Zhen-Yu Zhu, En-Ping Zhou, and Ang Li,  {\it Astrophysical Journal}, {\bf 862}, 98 (2018).
\bibitem{rn18} Rana Nandi and Prasanta Char,  {\it Astrophysical Journal}, {\bf 857} 12,  (2018).
\bibitem{gb19} Gordon Baym and et al, {\it Astrophysical Journal}, {\bf 885} 42, ( 2019).
\bibitem{da21} Daniel A. Godzieba, David Radice, and Sebastiano Bernuzzi, {\it Astrophysical Journal} {\bf 908}, 122 (2021).
\bibitem{gb05} G. Bertone, D. Hooper, and J. Silk, {\it Physics Reports} {\bf 405}, 279 (2005).
\bibitem{ad14}  A. Del Popolo, {\it Int. J. of Mod. Phys. D} {\bf 23}, 231430005 (2014).

\bibitem{pc16} Planck Collaboration, P. A. R. Ade, N. Aghanim, M. Arnaud  et al., {\it A \& A} {\bf 594}, 13 (2016).

\bibitem{ag01} A. Grant et al, {\it Astrophys. J.} {\bf 560}, 49 (2001). 
\bibitem{sp199} S. Perlmutter, M. S. Turner and M. White, {\it Phys. Rev. Lett.} {\bf 83}, 670 (1999).
\bibitem{cl03} C. L. Bennett et al, {\it Astrophys. J. Suppl.} {\bf 148}, 1 (2003).
\bibitem{gh03} G. Hinshaw et al. [WMAP Collaboration], {\it Astrophys. J. Suppl.} {\bf 148}, 135 (2003).
\bibitem{na03} N. Aghanim and Et al., {\it Astronomy \& Astrophysics} {\bf 641}, 6 (2020). 
\bibitem{pj03} P. J. E. Peebles and B. Ratra, {\it Reviews of Modern Physics} {\bf 75}, 559 (2003). 
\bibitem{ej06} E. J. Copeland, M. Sami and S. Tsujikawa, {\it Int. J. of Mod. Phys. D} {\bf 15}, 1753 (2006). 
\bibitem{rr09} R. R. Caldwell and M. Kamionkowski, {\it Annual Review of Nuclear and Particle Science} {\bf 59}, 397 (2009). 
\bibitem{dn15} D. N. Spergel, {\it Science (New York, N.Y.)} {\bf 347}, 1100 (2015).
\bibitem{bw16} B. Wang, E. Abdalla, F. Atrio-Barandela and D. Pavon, {\it Reports on progress in physics}, {\it Physical Society (Great Britain)} {\bf 79} (2016).
\bibitem{na20} N. Aghanim and Et al., {\it Astronomy \& Astrophysics} {\bf 641}, 1 (2020).
\bibitem{aa22} A. Addazi, {\it Physics of the Dark Universe} {\bf 37},  101102 (2022).
\bibitem{gb22} G. Bargiacchi et al., {\it Monthly Notices of the Royal Astronomical Society} {\bf 515}, 1795-1806 (2022).
\bibitem{sc22} S. Capozziello, D. Rocco, and L. Orlando,  {\it Physics of the Dark Universe} {\bf 36}, 101045 (2022).
\bibitem{os22} O. Sokoliuk, B. Alexander, and P. K. Sahoo,  {\it Physics Letters B} {\bf 829}, 137048 (2022).

\bibitem{CG92} G. Chapline, in Foundations of Quantum Mechanics, ed. by T. D. Black et al. (World Sci., Singapore, 1992).

\bibitem{JB04} J. Barbierii and G. Chapline, {\it Phys Lett. B} {\bf 590}, 8 (2004).

\bibitem{gc01} G. Chapline, E. Hohlfeld, R. B. Laughlin and D. Santiago, {\it Phil. Mag. B} {\bf 81}, 235 (2001).
\bibitem{sn06} S. N. F. Lobo, {\it Class. Quantum. Grav.} {\bf 23}, 1525 (2006).

\bibitem{lh20} L. Herrera,  {\it Phys. Rev. D} {\bf 101}, 104024 {2020}.
\bibitem{kg22} K. G. Sagar, B. Pandey, and N. Pant, {\it Astrophysics and Space Science}, {\bf 367}, 72 (2022).
\bibitem{ss22} S. Saklany, B. Pandey, and N. Pant,  {\it Modern Physics Letters A} {\bf 37}, 2250182 (2022). 
\bibitem{kg22a} K. G. Sagar, N. Pant, and B. Pandey,  {\it Physics of the Dark Universe} {\bf 38}, 101125 (2022).

\bibitem{pb15} P. Bhar and F. Rahaman, {\it Eur. Phys. J. C} {\bf 75}, 41 (2015).
\bibitem{ak11} A. K. Yadav, F. Rahaman and S. Ray, {\it Int. J. Theor. Phys.} {\bf 50}, 871 (2011).
\bibitem{ss11} S. S. Yazadjiev, {\it Phys. Rev. D} {\bf 83}, 127501 (2011).
\bibitem{bd21} B. Dayanandan and T. T. Smitha, {\it Chin. J. Phys.} {\bf 71}, 683 (2021).
\bibitem{ss21} S. Smerechynskyi, M. Tsizh, and B. Novosyadlyj, {\it JCAP} {\bf 02}, 045 (2021).
\bibitem{PR24} P. Rej, R. S. Bogadi, and M. Govender, {\it Chinese Journal of Physics} {\bf 87}, 608 (2024).
\bibitem{PB21} P. Bhar, {\it Physics of the Dark Universe} {\bf 34} 100879 (2021).
\bibitem{PB18} P. Bhar, T. Manna, F. Rahaman, and A. Banerjee, {\it Can. J. Phys.} {\bf 96}, 594 (2018).
\bibitem{RB16} R. Bibi, T. Feroze, and A. A. Siddiqui, {\it Can. J. Phys.} {\bf 94}, 758 (2016).
\bibitem{KD15} K. Das and N. Ali, {\it Astrophys. Space Sci.} {\bf 356}, 57 (2015).
\bibitem{PH13} P. Halpern and M. Pecorino, {\it ISRN Astron. Astrophys.} {\bf 2013}, 939876 (2013).
\bibitem{FR12} F. Rahaman Rahaman, R. Maulick, Y. K. Yadav, S. Ray, Saibal and R. Sharma, {\it General Relativity and Gravitation} {\bf 44}, 107-124 (2012).
\bibitem{AB22} A. B. Tudeshki, G. H. Bordbar, and B. E. Panah, {\it  Physics Letters B} {\bf 835}, 137523 (2022).
\bibitem{KP23} K. P. Das,  U. Debnath, and S.  Ray,  {\it Fortschritte der Physik} {\bf 71}, 2200148 (2023).
\bibitem{RC09} R. Chan, M. F. A. Da Silva, and J. F. V. da Rocha, {\it General Relativity and Gravitation} {\bf 41}, 1835-1851 (2009).
\bibitem{AB23} A. B. Tudeshki, G. H. Bordbar, and B. E. Panah, {\it Physics of the Dark Universe} {\bf 42}, 101354 (2023).
\bibitem{CR11} C. R. Ghezzi, {\it Astrophysics and Space Science} {\bf 333} 437-447 (2011).
\bibitem{SS12} S. S. Yazadjiev, and D. D. Doneva, {|it JCAP}  {\bf 2012}, 037 (2012).
\bibitem{DH13} D. Horvat, and A. Marunović,  {\it Classical and quantum gravity} {\bf 30}, 145006 (2013).
\bibitem{PR23} P. Rej, A. Karmakar, {\it Eur. Phys. J. C}  {\bf 83}, 699 (2023).

\bibitem{GC67} G. Chodil, H. Mark, R. Rodrigues, F. D. Seward, C. D. Swift,  {\it Astrophysical Journal} {\bf 150}, 57 (1967).
\bibitem{RG72} R. Giacconi, et al., {\it Astrophysical Journal} {\bf 178}, 281 (1972).
\bibitem{RG71} R. Giacconi, H. Gursky, E. Kellog, E. Schreier, H. Tananbaum, {\it Astrophysical Journal} {\bf 167}, 67 (1971).
\bibitem{ES72}  E. Schreie, R. Levinson, H. Gursky, E. Kellog, H. Tananbaum, R. Giacconi, {\it Astrophysical Journal}, {\bf 172}, 79 (1972).
\bibitem{RE71} R. E. Price, D. J. Groves, R. M. Rodrigues et al. {\it ApJL} {\bf 168} L7 (1971).
\bibitem{CL71} C. Leong , E. Kellogg, H. Gursky, H. Tananbaum and R. Giacconi, {\it ApJL} {\bf 170} L67 (1971).
\bibitem{ES72a} E. Schreier, R. Giacconi, H. Gursky, E. Kellogg and H. Tananbaum, {\it ApJL} {\it 178} {\bf L71} (1972).
\bibitem{CC77} C. Chevalier, S. A. Ilovaisky, {\it A \& A} {\bf 59}, L9 (1977).
\bibitem{cr11} C. R. Ghezzi, {\it Astrophys. Space Sci.} {\bf 333}, 437 (2011).
\bibitem{hh69} H. Heintzmann,  {\it Zeitschrift f¨ur Physik } {\bf 228}, 489-493 (1969).
\bibitem{cr05} C. R. Ghezzi, {\it Phys. Rev. D} {\bf 72}, 104017 (2005).
\bibitem{wb07} W. Barreto, B. Rodriguez, L. Rosales, O. Serrano, {\it Gen. Rel. Grav.} {\bf 39}, 23 (2007).
\bibitem{nk12} N. K. Glendenning, Compact Stars: Nuclear Physics, Particle Physics and General Relativity (Springer Science \& Business Media, Berlin, 2012).
\bibitem{cs22} C. S. Chu, H.S. Tan, {\it Universe} {\bf 8}, 250 (2022).
\bibitem{gd27} G. Darmois, Les equations de la gravitation einsteinienne. Mémorial des Sciences Mathematiques, no 27, p. 58 (1927).
\bibitem{wi67} W. Israel, Nuovo Cim. B 44S10, 1 (1966) [Erratum: Nuovo Cim.B
48, 463 (1967)]
\bibitem{sc84} S. Chandrasekhar, {\it Science} {\bf 226}, 497 (1984).
\bibitem{lb07} L. Baum, P. H. Frampton, {\it Phys. Rev. Lett.} {\bf 98}, 071301 (2007).
\bibitem{ha59} H. A. Buchdahl, {\it Phys. Rev.} {\bf 116}, 1027 (1959).
\bibitem {step} S. W. Hawking, G.F.R. Ellis, {\it The Large Scale Structure of Space-Time}, (Cambridge University Press, Cambridge, 1973).
\bibitem {wald} R. M. Wald, {\it General Relativity} (University of Chicago Press, Chicago, 1984).
\bibitem{sl08} S. L. Shapiro, S.A. Teukolsky, Black Holes, White Dwarfs, and Neutron Stars: The Physics Of Compact Objects (Wiley, New York, 2008).
\bibitem{yb71} Y. B. Zeldovich, I. D. Novikov, {\it Relativistic Astrophysics Vol. 1: Stars and Relativity} (University of Chicago Press, Chicago, 1971).
\bibitem{yb72} Y. B. Zeldovich, I.D. Novikov, {\it J. Silk, Phys. Today} {\bf 25}, 63 (1972).
\bibitem{yl62} Y. L’Dovich, {\it Sov. Phys. JETP} {\bf 14}, 1609 (1962).
\bibitem{hb64} H. Bondi, {\it Proceedings of the Royal Society of London. Series A. Mathematical and Physical Sciences} {\bf 281}, 39–48 (1964).
\bibitem{hh75} H. Heintzmann and W Hillebrandt, {\it Astronomy and Astrophysics} {\bf 38}, 51–55, (1975).
\bibitem{rc93} R. Chan, L. Herrera, and N. O. Santos, {\it Monthly Notices of the Royal Astronomical Society} {\bf 265}, 533–544 (1993).
\bibitem{chc17} C. C. Moustakidis, {\it General Relativity and Gravitation} {\bf 49}, 68, (2017).

\bibitem {har65}  B. K. Harrison, K.S. Thorne, M. Wakano, J.A. Wheeler, Gravitational theory and gravitational collapse (University of Chicago Press, Chicago, 1965).
\bibitem {ml11}  M. L. Rawls, J. A. Orosz, J. E. McClintock, M. A. P. Torres, C. D. Bailyn, and M. M. Buxton, {\it Astrophys. J.} {\bf 730}, 25 (2011).
\bibitem {ab17} A. Bauswein et al., {\it Astrophys. J. Lett.} {\bf 850}, L34 (2017).
\bibitem {cd20}  C. D. Capano et al., {\it Nat. Astron.} {\bf 4}, 625 (2020).
\bibitem{ruff} C. E. Rhoades, R. Ruffini, {\it Phys. Rev. Lett.} {\bf 32}, 324 (1972).
\bibitem {latt} J. M. Lattimer, M. Prakash, {\it Phys. Rep.} {\bf 442}, 109 (2007).
\bibitem{hart} J. B. Hartle, {\it Astrophys. J.} {\bf 150}, 1005  (1967).
\bibitem{bejg} M. Bejger, P. Haensel, {\it A \& A} {\bf 396}, 917 (2002).
\bibitem{bv02} B. V. Ivanov, {\it Phys. Rev. D} {\bf 65}, 104011 (2002).

\end{thebibliography}
\end{document}